\documentclass[a4,12pt]{article}
\usepackage[dvips]{graphicx}
\usepackage{amsmath,amssymb}
\newcommand{\Mat}[1]{{{\boldsymbol{#1}}}}
\newcommand{\abs}[1]{\left\vert#1\right\vert}
\date{}
\begin{document}
\title{\bf Gravitational effects on light rays and binary pulsar energy loss in a scalar theory of gravity}
\author{M. Arminjon}
%
%

\maketitle
\begin{center}
{ \small Laboratoire ``Sols, Solides, Structures'' [Unit\'e Mixte de Recherche of the CNRS], 
\\ BP 53, F-38041 Grenoble cedex 9, France}
\\ email: arminjon@hmg.inpg.fr

\vspace{2cm}

\begin{abstract} A scalar bimetric theory of gravity with a preferred reference frame is summarized. Dynamics is governed by an extension of Newton's second law. In the static case, geodesic motion is recovered together with Newton's attraction field. In the static spherical case, Schwarzschild's metric is found. Asymptotic schemes of post-Newtonian (PN) and post-Minkowskian (PM) approximation are built, each based on associating a conceptual family of systems with the given system. At the 1PN approximation, there is no preferred-frame effect for photons, hence the standard predictions of GR for photons are got. At the 0PM approximation, an isolated system loses energy by quadrupole radiation, without any monopole or dipole term. Inserting this loss into the Newtonian 2-body problem gives the Peters-Mathews coefficients of the theory.
\end{abstract}

\end{center}
\section{Introduction} 

According to a common belief, a scalar theory should not be able to explain the effects of gravitation on light rays, which have been predicted on the basis of general relativity (GR), and have received experimental confirmation. It is true that the most obvious covariant scalar theory does not predict any light bending \cite{MTW}, for instance, but this particular example does not prove that {\it no} scalar theory can predict light bending. In the same vein, it is usually thought that a scalar theory should lead to monopole and dipole gravitational radiation, and that, therefore, the effects of gravitational radiation should be very important according to a such theory---contrary to observation. The latter point is believed to be even more true if a {\it preferred-frame} scalar theory is considered. For instance, Carlip \cite{Carlip} writes: ``For a [scalar] non-Lorentz invariant theory, conservation laws place no restrictions on radiation, and monopole radiation should dominate, permitting aberration at all orders.'' However, the latter statement is in the form of a conjecture, which has to be checked for each specific theory.\\

The aim of this paper is: {\bf i}) to present to the readers of this Journal a summary of the scalar theory with a preferred reference frame \cite{A18, A15, A20} and of the proof \cite{A19} that this theory predicts just the standard effects of GR on light rays; {\bf ii}) for the first time, to investigate gravitational radiative energy loss in this scalar theory. In particular, it will be shown that this is indeed a {\it loss}, and that there is neither monopole nor dipole gravitational radiation for this theory.

\section{Summary of the scalar ether-theory} \label{summary}

The starting point of the theory is the {\it heuristic interpretation} of gravity as resulting from the pressure force exerted on matter at the scale of elementary particles by an imagined perfect fluid or ``micro-ether''. This interpretation leads to a definite {\it set of equations}: it is this set that really constitutes the theory, and that should be assessed from the predictions it leads to and the comparison of these predictions with experimental data. According to this heuristic view, gravity would be simply ``Archimedes' thrust in the ether''; it turns out that Newtonian gravity had been interpreted in this way by Euler in 1746 \cite{Euler}. The construction of the theory along this line is given in Ref. \cite{A18}, but the formulation of motion was later modified in order to obtain a conservation equation for the energy \cite{A15}. \footnote{The publication dates are confusing here: the first version of Ref. \cite{A18} dates back to end 1992.} Since then, the general motion of a test particle is governed by an extension of Newton's second law (formerly \cite{A18}, this was true only in the static case); this applies also to light-like particles (photons) \cite{A19}. The dynamical equations that apply to a continuous medium have been deduced from the dynamics of a test particle, by considering first a dust continuum \cite{A20}. These dynamical equations allow to define the gravitational extension of the Maxwell equations in the theory, which is consistent with photon dynamics as governed by the extension of Newton's second law (\cite{B13}, part 3).

\subsection{Space-time metric and field equation} \label{metric+field}

The equations are given in the {\it preferred reference frame} $\mathrm{E}$, which is the rest frame of the reference body $\mathrm{M}$  or ``macro-ether''. 
\footnote{
A reference frame can be physically thought as a spatial network of observers equipped with measuring rods and clocks, allowing to define space-time coordinates $(z^\mu)$ ($\mu = 0,...,3$) such that each observer has constant space coordinates $z^i$  ($i = 1,2,3$). Such coordinates are called ``adapted to the frame'', see Cattaneo \cite{Cattaneo}. The 3-D reference body associated with a frame may be defined as the set of the world-lines of the observers, and may be described by the space coordinates $(z^i)$.
}
The reference body $\mathrm{M}$ is assumed to be equipped with an Euclidean space metric $\Mat{g}^0$. Observers bound to the frame $\mathrm{E}$ would measure the spatial distances defined with $\Mat{g}^0$, and would also measure the ``absolute time'' $T$, if there were no gravitation. However, an ``absolute'' version of Einstein's equivalence principle implies that physical standards of space (resp. time) are contracted (resp. dilated) in a gravitational field \cite{A18}. The scalar field of the theory is the contraction factor $\beta$ affecting the time interval $dt_\mathbf{x}$ measured by a clock fixed at point $\mathbf{x}$  bound to the frame $\mathrm{E}$ (due to the dilation of the clock period in the gravitational field), or the square $f = \beta^2$. The latter defines the $\gamma_{00}$ component of the physical, curved space-time metric $\Mat{\gamma}$, in any coordinates $(x^\mu)$, with $x^0 = cT$ (and $c$ the velocity of light), that are bound to the frame $\mathrm{E}$:

\begin{equation} \label {localtime}
		 			dt_\mathbf{x}/dT = \surd f \equiv \beta \leq 1	.
\end{equation}
The interpretation of gravity as Archimedes' thrust leads to a definite expression of the acceleration vector $\mathbf{g}$ of gravitation, in terms of the gradient of the ``ether pressure'' $p_e$, the gradient operator being defined with the physical space metric $\Mat{g}$ \cite{A18}. It turns out \cite{A15} that $\mathbf{g}$ may be reexpressed in a very simple form in terms of the Euclidean metric $\Mat{g}^0$ and the scalar $f$:

\begin{equation} \label {g_grav}
		 			\mathbf{g}=-\frac{c^2}{2}\nabla_0f, \quad \nabla_0f\equiv \mathrm{grad}_{\Mat{g}^0}f, \quad (\mathrm{grad}_{\Mat{g}^0}f)^i\equiv g^{0ij}f_{,j}.
\end{equation}
(Matrix $(g^{0ij})$ is the inverse of the component matrix $(g_{0ij})$ of the Euclidean space metric $\Mat{g}^0$.) Due to the ``absolute'' version of the equivalence principle, the field $\beta$ defines also the dilation of an infinitesimal distance $dl$ measured in the direction of vector $\mathbf{g}$, as compared with the distance $dl^0$ evaluated using the Euclidean metric: $dl = dl^0/\beta$, while the infinitesimal distances in directions perpendicular to $\mathbf{g}$ are unaffected. This assumption may be reexpressed in saying that the physical space metric $\Mat{g}$ in the frame $\mathrm{E}$ has the following relation to the Euclidean metric $\Mat{g}^0$:
\footnote{
We note that Eq.~(\ref{spacemetric}) does not apply when $ \mathbf{g} =\mathbf{0}$. Indeed, the assumed contraction occurs in the direction of the gravity acceleration $\mathbf{g}$, hence it is {\it a priori} undefined at those places where $ \mathbf{g} =\mathbf{0}$ [unless the scalar potential $f $ has the same value in that place as the one it has ``at infinity'' ({\it i.e.}, $f = 1$): then, there is no contraction at all]. In the generic, physically realistic case, the vector $\mathbf{g}$ vanishes merely at isolated points, e.g. at one point between two massive bodies, and also at one point near the centre of each massive body. At an isolated point $\mathbf{x}$ such that $\mathbf{g}(\mathbf{x}) = \mathbf{0}$, the direction of contraction cannot be defined uniquely. However, Eq.~(\ref{spacemetric}) applies in the neighborhood of $\mathbf{x}$ and shows that the metric $\Mat{g}$ is discontinuous at this isolated point, but remains bounded in its neighborhood. Hence, geometrical elements such as the length of a line are defined as with a continuous metric. Moreover, the examination of the case with spherical symmetry shows that no difficulty occurs in the dynamical equations either. A more difficult case occurs in the highly idealized situation of ``collapse in free fall'': $\mathbf{g}$ vanishes in a bounded volume domain \cite{A18}. This case was easily solved in Ref. \cite{A18} for the relevant subcase of spherical symmetry, and the way of solution can be extended to the case without any symmetry. From a theoretical point of view, one should remind that, in this theory, the ``true'' geometry is in terms of the Euclidean metric $\Mat{g}^0$, whereas $\Mat{g}$ has merely the status of an ``effective metric''---as is also the case in the ``relativistic theory of gravitation'' (RTG) \cite{Logunov88, Logunov89}.
}

\begin{equation} \label {spacemetric}
		 			\Mat{g} =	\Mat{g}^0 +\left(\frac{1}{f}-1\right)\Mat{h}, \quad\Mat{h}\equiv\frac{\mathbf{g}^*\otimes \mathbf{g}^*} {\mathbf{g}^2}      
\end{equation}
(with $\mathbf{g}^*\equiv (\mathsf{g}_i), \quad\mathsf{g}_i\equiv g^0_{ij}\mathsf{g}^j$ and ${\mathbf{g}^2}\equiv g^0_{ij}\mathsf{g}^i  \mathsf{g}^j $). The space-time metric writes, in any coordinates $(x^\mu)$ bound to the frame $\mathrm{E}$, with $x^0 = cT$:

\begin{equation} \label {spacetimemetric}
ds^2 = \gamma_{\mu\nu} dx^\mu dx^\nu = f (dx^0)^2 - dl^2,  dl^2 = g_{ij} dx^i dx^j,
\end{equation}
that is, $\gamma_{00} = f,   \gamma_{ij} = - g_{ij},    \gamma_{0i} = 0$. That the $\gamma_{0i}$ components are always zero in the ether frame $\mathrm{E}$, is the natural translation of the assumption that this frame admits a global simultaneity defined with the absolute time $T$. \\

In many problems, one may neglect the time variation of the ``reference ether pressure'' \cite{A18}, $p_e^\infty (T)$, because the analysis of cosmological models \cite {A28} shows that this variation takes place over long time scales, of the order of $10^8$ years. Then the field equation \cite{A18} may be rewritten \cite{A15} as:

\begin{equation} \label{field}
\Delta f- \frac{1}{f} \left(\frac{f_{,0}}{f}\right)_{,0}= \frac{8\pi G}{c^2} \sigma \qquad (x^0 = cT),
\end{equation}
with $\Delta \equiv \mathrm{div}_{\Mat{g}^0} \mathrm{grad}_{\Mat{g}^0}$ the usual Laplace operator, defined with the Euclidean metric $\Mat{g}^0$, and where $G$ is Newton's gravitation constant, and $\sigma \equiv (T^{00})_{\mathrm{E}}$ is the mass-energy density in the ether frame. ($\mathbf{T}$ is the ``mass tensor'', {\it i.e.} the energy-momentum tensor in mass units.) This equation is merely space-covariant. It is immediate that, in any static situation, Eqs. (\ref{g_grav}) and (\ref{field}) associate just the {\it Newtonian} gravity acceleration with the given constant mass-energy density $\sigma = \sigma(\mathbf{x})$. However, $\sigma  \equiv T^{00}$ does not only involve the density of the pure matter energy (e.g., for ``dust'', the latter would be the rest-mass plus kinetic energy), for it is also increased by the gravitational field itself \cite {A15}. In the static spherically symmetric situation, the space-time metric (\ref{spacetimemetric}) predicted from the field equation (\ref{field}) and the boundary conditions ($f(r)\rightarrow 1$ as $r\rightarrow \infty $ and $f(r)$ bounded as $r\rightarrow 0$) is quickly got in spherical coordinates $r$, $\theta$ and $\varphi$ for $\Mat{g}^0$:

\begin{equation} \label{SSSmetric}
ds^2 = \left (1- \frac{2V(r)}{c^2} \right) c^2 \,dT^2 - \frac{1}{1- \frac{2V(r)}{c^2}}\,dr^2 - r^2 \,d\Omega ^2, \   d\Omega ^2 \equiv d\theta ^2 + \sin^2 \theta \,d\varphi ^2, 
\end{equation}
where $V(r)$ is the Newtonian potential associated with the density $\sigma = \sigma(r)$:

\begin{equation} \label{SSSpotential}
V(r) = \int_{r}^{\infty} \frac{GM(u)}{u^2}\,du,\quad M(r) \equiv 4\pi\int_{0}^{r} u^2 \sigma (u) du. 
\end{equation}
($M \equiv M(\infty) \equiv \lim_{\,r\rightarrow \infty} M(r)$ is assumed finite.) Being a static solution, this solution assumes that the mass-energy density $\sigma$ is time-independent (and spherically symmetric) {\it in the preferred frame}, {\it i.e.}, the massive body is at rest in the ``ether'' or in a stationary rotation with respect to it. If $\sigma(r) = 0$ for $r > R$, then $V(r) = GM/r$ for $r > R$, this is the {\it Schwarzschild exterior solution}. As to the Schwarzschild interior solution of GR, it is only known when, in addition, $\sigma (r) = \mathrm{Const}$. for $0 \leq r < R$, but Eq.~(\ref{SSSmetric}) does {\it not} give the Schwarzschild interior solution in that case. Note that, in contrast with the situation in GR, Eq.~(\ref{SSSmetric}) is valid as soon as $M(\infty) < \infty$ and makes no difference between the ``interior'' and ``exterior'' cases.

\subsection{Dynamical equations} \label{dynamics}
\subsubsection{Motion of a test particle} \label{testparticle}
This is governed by a natural extension of Newton's second law, the latter being defined in terms of the {\it physical}, curved metric. In the case of a constant gravitational field, it is relatively straightforward to extend Planck's special-relativistic form of Newton's second law ($\mathbf{F} = d\mathbf{P}/dt$ with $\mathbf{F}$ the force and $\mathbf{P}$ the momentum) to the situation with gravitation, because the time-derivative of a vector (like $\mathbf{P}$) must undoubtedly be defined as the ``absolute derivative'', the latter being induced from the covariant derivative with respect to the constant space metric $\Mat{g}$. In coordinates, the absolute derivative of a vector $\mathbf{w}$ depending on the parameter $\xi$ along a trajectory in a Riemannian space equipped with a fixed metric $\Mat{g}$ is given by:

\begin{equation} \label{absolutederivative}
\left(\frac{D_0\mathbf{w}}{D\xi}\right)^i \equiv \frac{dw^i}{d\xi}\,+ \Gamma^i_{jk} w^j\,\frac{dx^k}{d\xi},
\end{equation}
with $\Gamma^i_{jk}$ the Christoffel symbols of metric $\Mat{g}$. This derivative was used by Landau \& Lifshitz \cite{L&L} to rewrite the spatial part of Einstein's geodesic equation of motion as Newton's second law, for a constant gravitational field. In the case of a variable gravitational field, we still have, in the frame $\mathrm{E}$ (as also in any other possible reference frame), a well-defined space metric $\Mat{g}$---but now it depends on time, and we have a local time $t_{\mathbf{x}}$. We have found \cite{A16} that a unique definition may be given for the time-derivative $D\mathbf{w}/D\xi$ of a space vector $\mathbf{w}$ depending on the parameter $\xi$ along a trajectory in the given space $\mathrm{M}$ endowed with a metric $\Mat{g} = \Mat{g}_\xi$ that varies with parameter $\xi$, under the following essential requirements: (i) it must be a space vector depending linearly on $\mathbf{w}$, (ii) it must reduce to the absolute derivative $D_0\mathbf{w}/D\xi$ if $\Mat{g}$ does not actually depend on $\xi$, (iii) it must be expressed in terms of the space metric $\Mat{g}$ and its derivatives, and (iv) it must obey Leibniz' derivation rule for a scalar product, {\it i.e.}

\begin{equation} \label{Leibniz}
\frac{d}{d\xi}\left(\Mat{g}(\mathbf{w}, \mathbf{z})\right) = \Mat{g}\left(\mathbf{w}, \frac{D\mathbf{z}}{D\xi}\right) + \Mat{g}\left(\frac{D\mathbf{w}}{D\xi},\mathbf{z}\right). 
\end{equation}
This unique definition is as follows \cite{A16}:

\begin{equation} \label{vectorderivative}
\frac{D\mathbf{w}}{D\xi} \equiv \frac{D_0\mathbf{w}}{D\xi} \,+ \frac{1}{2}\Mat{t}\mathbf{.w},\quad \Mat{t} \equiv \Mat{g}_\xi ^{-1}\mathbf{.}\frac{\partial \Mat{g}_\xi } {\partial \xi}. 
\end{equation}
To define Newton's second law, the trajectory of the test particle is parametrized with the local time: $\xi = t_{\mathbf{x}}$, synchronized along the trajectory as defined by Landau \& Lifshitz \cite{L&L}. Actually, definition (\ref{vectorderivative}) is also valid (and compelling under the above requirements) for any reference frame in any theory with a curved space-time metric \cite{A16}. In the present case, the synchronized local time is simply obtained by integration of Eq.~(\ref{localtime}) along the trajectory, because the absolute time $T$ is globally synchronized in the ether frame $\mathrm{E}$  ({\it i.e.}, because the $\gamma_{0i}$ components are assumed zero in $\mathrm{E}$). 

There is no ambiguity for the definition of the momentum, either. This is

\begin{equation} \label{momentum}
\mathbf{P} \equiv m(v) \mathbf{v}, \quad m(v) \equiv m(v = 0).\gamma_v \equiv m(0).(1-v^2/c^2)^{-1/2},		
\end{equation}
the velocity $\mathbf{v}$ of the test particle (relative to the frame $\mathrm{E}$) being measured with the local time $t_\mathbf{x}$, and its modulus $v$ being defined with the space metric $\Mat{g}$:

\begin{equation} \label{velocity}
v^i \equiv \frac{dx^i}{dt_\mathbf{x}} \equiv \frac{1}{\surd f} \frac{dx^i}{dT}, \quad v \equiv \Mat{g}(\mathbf{v},\mathbf{v})^{1/2} = (g_{ij} v^i v^j)^{1/2}.	
\end{equation}		
The gravitational force is assumed to be $\mathbf{F}_\mathrm{g} = m(v)\mathbf{g}$ with $\mathbf{g}$ given by Eq.~(\ref{g_grav}), and there may be also a non-gravitational force $\mathbf{F}_0$. Thus, the extension of Newton's second law is obtained as

\begin{equation} \label{Newtonlawmasspoint} 
\mathbf{F}_0 + m(v)\mathbf{g} = \frac{D\mathbf{P}}{Dt_\mathbf{x}}.		 		 
\end{equation}

For a photon, the velocity $\mathbf{v}$ is still defined by (\ref{velocity}), but the mass content of the energy $E = h\nu$ is substituted for the inertial mass, the frequency $\nu$ (number $n$ of oscillations per unit time) being measured with the local time:

\begin{equation} \label{energyphoton}
E \equiv h\nu,\quad \nu \equiv \frac{dn}{dt_{\mathbf{x}}},
\end{equation}

\begin{equation} \label{Newtonlaw}	                               
\mathbf{F}_0 + \frac{E}{c^2}\,\mathbf{g} = \frac{D\mathbf{P}}{Dt_\mathbf{x}},\quad \mathbf{P} \equiv \frac{E}{c^2}\,\mathbf{v}.             
\end{equation}

Defining the ``pure energy'' of a mass point ({\it i.e.}, not accounting for the potential energy in the gravitational field) as
 
\begin{equation} \label{energymasspoint}
E = m(v) c^2, 
\end{equation}
we see that Eq. (\ref{Newtonlaw}) applies to both mass points and photons. Together with the assumed vector $\mathbf{g}$ (Eq.~(\ref{g_grav})) and the assumed metric (Eqs.~(\ref{spacemetric}) and (\ref{spacetimemetric})), Eq. (\ref{Newtonlaw}) implies that free particles ($\mathbf{F}_0 = \mathbf{0}$) follow space-time geodesics in a {\it static} gravitational field ($f_{,0} = 0$): see Ref.~\cite{A16} for mass points and see Ref.~\cite{A18} for photons. Also, Eq. (\ref{Newtonlaw}) with $\mathbf{F}_0 = \mathbf{0}$ implies \cite{A15} a simple evolution equation for the energy $E$:
\begin{equation} \label{energyevolution}
\frac{d(E\beta)}{dT} = E \frac{\partial\beta}{\partial T} \quad (\beta \equiv \surd f).
\end{equation}
Using Eq. (\ref{energyevolution}), Newton's second law (\ref{Newtonlaw}) can be recast \cite{A19} to give the ``coordinate acceleration'' for a free test particle (mass point or photon):
\begin{equation} \label{coordinateacceleration}	                               
\frac{du^i}{dT}= \frac{1}{\beta}\left(\frac{\partial\beta}{\partial T} + 2\beta_{,j}u^j\right)u^i - \Gamma^i_{jk} u^j u^k - \frac{1}{2}g^{ij}\,\frac{\partial g_{jk}}{\partial T}u^k - \frac{c^2}{2}\,f(\nabla_0 f)^i.           
\end{equation}

\subsubsection{Continuum dynamics and energy conservation} \label{continuumdynamics}
Equations (\ref{vectorderivative}) (with $\xi = t_{\mathbf{x}}$) and (\ref{Newtonlaw}) (with $\mathbf{F}_0 = \mathbf{0}$) define Newton's second law in the preferred reference frame, for any free test particle. It is thereby defined also for dust, which is a continuum made of coherently moving, non-interacting particles, each of which conserves its rest mass. Dust is characterized by its mass tensor
\begin{equation} \label{dust}
T^{\mu \nu} = \rho^*U^\mu U^\nu 
\end{equation}
(with $\rho^*$ the proper rest-mass density and $U^{\mu} \equiv dx^{\mu}/ds$ the 4-velocity). Independently of the assumed form for the space-time metric $\Mat{\gamma}$ (provided it satisfies $\gamma_{0i} = 0$ in the preferred frame), this extension of Newton's second law mathematically implies \cite{A20} the following dynamical equation for dust:

\begin{equation} \label{continuum}
T_{\mu;\nu}^{\nu} = b_{\mu},				          
\end{equation}
where $b_\mu$ is defined by

\begin{equation} \label{definition_b}
b_0(\mathbf{T}) \equiv \frac{1}{2}\,g_{jk,0}\,T^{jk}, \quad b_i(\mathbf{T}) \equiv -\frac{1}{2}\,g_{ik,0}\,T^{0k}. 
\end{equation}
(Indices are raised and lowered with metric $\Mat{\gamma}$, unless mentioned otherwise. Semicolon means covariant differentiation using the Christoffel connection associated with metric $\Mat{\gamma}$.) Equation (\ref{continuum}), with the definition (\ref{definition_b}), is assumed to hold true for any material continuum: accounting for the mass-energy equivalence, this is the expression of the universality of gravitation. Equation (\ref{continuum}) is valid in any coordinates $(x^\mu)$ that are adapted to the frame $\mathrm{E}$ and such that $x^0 = \phi(T)$ with $T$ the absolute time. (In contrast, the validity of the scalar field equation (\ref{field}) demands that $x^0 = cT$.) For dust, Eq.~(\ref{continuum}) {\it implies}, conversely \cite{A20}, that the constitutive particles obey Newton's second law, and that the rest mass is conserved, {\it i.e.}
\begin{equation} \label{massconservation}
\left(\rho^*U^\nu\right)_{;\nu} = 0. 
\end{equation}

For a general continuum, Eq.~(\ref{continuum}) implies an exact energy conservation, which {\it substitutes} \cite{A20} for the exact mass conservation~(\ref{massconservation}). With the help of the following relation, valid for the assumed space-time metric (Eqs.~(\ref{spacemetric}) and~(\ref{spacetimemetric})):
\begin{equation} \label{determinant}
-\gamma = \gamma_{00}.g = f.(g^0/f) = g^0 \quad (\gamma \equiv \mathrm{det}(\gamma_{\mu\nu}),\ g \equiv \mathrm{det}(g_{ij}),\ g^0 \equiv \mathrm{det}(g^0_{ij})),
\end{equation}
one obtains from~(\ref{continuum}), in Cartesian coordinates $(x^i)$ ({\it i.e.}, $g^0_{ij} = \delta_{ij}$) and with $x^0 = cT$:
\begin{equation} \label{flat_0}
T_{0,\nu}^\nu =  \frac{1}{2}f_{,0}\,T^{00} \equiv \frac{1}{2} f_{,0}\,\sigma,
\end{equation}
\begin{equation} \label{flat_i}
T_{i,\nu}^\nu = \frac{1}{2} f_{,i}\,\sigma - \frac{1}{2}\,g_{ik,0}\,T^{0k} - \frac{1}{2}\,g_{jk,i}\,T^{jk}.
\end{equation}
Using the scalar field equation (\ref{field}), Eq.~(\ref{flat_0}) may then be rewritten as \cite{A15}
\begin{equation} \label{energyconservation}
T_{0,\nu}^\nu + \frac{1}{8\pi G}\left\{ 
\left[\frac{\mathbf{g}^2}{c^2} + \frac{c^2}{4}\left(\frac{f_{,0}}{f}\right)^2\right]_{,0} 
+ \mathrm{div}_{\Mat{g}^0}\left(f_{,0}\mathbf{g}\right)\right\} = 0.
\end{equation}
This is a true conservation equation for the quantity
\begin{equation} \label{energydensity}
\varepsilon \equiv c^2 T_0^0 + \frac{1}{8\pi G} \left[\mathbf{g}^2 + \frac{c^4}{4}\left(\frac{f_{,0}}{f}\right)^2\right] \equiv \varepsilon_\mathrm{m} + \varepsilon_\mathrm{g}, 
\end{equation}
which may be called the total energy density, while $\varepsilon_\mathrm{m} \equiv c^2T_0^0$ and $\varepsilon_\mathrm{g}$ are the volume densities of the matter energy and the pure gravitational energy. Integrating (\ref{energyconservation}) in a fixed domain $\Omega$, and assuming that there is no flux of matter accross its boundary $\partial \Omega$, we get
\begin{equation} \label{energyvariation}
\frac{dE}{dT} = -\Phi , \quad \Phi \equiv \frac{c^2}{8\pi G} \int_{\partial \Omega} \frac{\partial f}{\partial T}\,\mathbf{g.n}\,dS, \quad E \equiv \int_{\Omega} \varepsilon\,dV
\end{equation}
($V$ and $S$ are the volume and surface measures associated with the Euclidean spatial metric $\Mat{g}^0$; in (\ref{energyconservation}-\ref{energyvariation}) the scalar product and square also refer to $\Mat{g}^0$).

\section{Effects of gravitation on light rays}
The prediction of gravitational effects on light rays in GR is done by assuming that the world lines of the light rays are null geodesics of the space-time metric, and that there is just one massive body, which is static and spherically symmetric (see {\it e.g.} refs. \cite{MTW, Weinberg, Will}). This seems justified: in view of the ({\it a posteriori}) smallness of these effects, the deviation from a uniform motion becomes sensible only when the ray becomes very close to one massive body, so that the field of the other bodies may indeed be neglected; and the small deviation from sphericity is a second-order effect. As we have recalled above, in the static situation, the extension (\ref{Newtonlaw}) of Newton's second law assumed in the present theory implies \cite{A18} that photons follow space-time geodesics of metric $\Mat{\gamma}$, and these are indeed null geodesics, since the velocity $v$ of the photons (as defined by Eq.~(\ref{velocity})) is the constant $c$. 
\footnote{
In Ref. \cite{A18}, photons are {\it defined} as light-like particles in the sense that $v = c$, in the same way as in GR. (Of course, $v = c$ is equivalent to say that the tangent vector to the world line of the particle, $\mathbf{U} = (U^\mu)$ with $U^\mu \equiv dx^\mu/d\xi$, is an ``isotropic'' 4-vector, {\it i.e.} $\Mat{\gamma}(\mathbf{U},\mathbf{U})=0$.) The assumption $v = c$ is compatible with Newton's second law (\ref{Newtonlaw}) with $\mathbf{F}_0 =\mathbf{0}$ . Indeed, the coordinate acceleration (\ref{coordinateacceleration}) can be rewritten to give
\begin{equation}
\frac{D\mathbf{v}}{DT}=\surd f \left(\mathbf{g}-(\mathbf{g.v})\frac{\mathbf{v}}{c^2}\right), \quad\mathbf{v} = \frac{1}{\surd f} \frac{d\mathbf{x}}{dT} 
\end{equation}
(where $\mathbf{g.v}\equiv\Mat{g}(\mathbf{g,v})$). By calculating $d(v^2)/dT$, is easy to check that any solution of this with an initial data such that $v(T=0)=c$ satisfies $v=c$ at any time.
}
Moreover, in the static spherically symmetric situation, the metric $\Mat{\gamma}$ (\ref{SSSmetric}) predicted by the present theory outside the body is just Schwarzschild's metric. One thus expects that this theory should predict the same effects on light rays as the standard effects predicted with the Schwarzschild metric in GR. However, in this preferred-frame theory, we cannot assume that the relevant body is at rest in the preferred reference frame $\mathrm{E}$: we should account for a uniform translation of the body through this ``ether''. 
\footnote{
If the ``absolute'' velocity $v_{\mathrm{B}}$ of the body is negligible with respect to $c$, we may consider that $v_{\mathrm{B}}$ is indeed a constant, at least when calculating the trajectories of light rays. Now this theory predicts preferred-frame effects in celestial mechanics, and these have to be small, in view of the fact that Newtonian mechanics (which has no preferred-frame effect) is already very accurate. Hence it is necessary that $v_{\mathrm{B}}$ be indeed small. The adjustment of the equations of motion for the mass centers \cite{A32} on observations will allow to determine a ``best-fitting'' value of the absolute velocity $\mathbf{V}$ of the mass center of the solar system, hence giving access to absolute velocities of all solar-system bodies. A preliminary adjustment (neglecting the effects of self-rotation of the celestial bodies, and assuming that the solar system is isolated) has led to the value $\abs{\mathbf{V}}\simeq 3$ km/s \cite{B21}. Hence one expects that the absolute velocities of solar-system bodies should have the same order of magnitude as their orbital velocities and, in particular, should indeed be negligible with respect to $c$.
}
\subsection{Asymptotic post-Newtonian approximation} \label{PNA}
To do this, we must develop an approximation scheme valid for weak gravitational fields and for low velocities of celestial bodies, thus a ``post-Newtonian'' (PN) approximation scheme. This should be done in following as closely as possible the rigorous procedures which are currently used in applied mathematics, because relativistic celestial mechanics aims at very high precisions. A PN approximation has been initiated along this line for GR by Futamase \& Schutz \cite{FutaSchutz}, and another one, which bears several similarities with the former, has been (independently) proposed and developed for the scalar ether-theory in Refs. \cite{A19, A23}. We call this scheme the ``asymptotic'' PN scheme, because it is based on the usual method of asymptotic expansion, as applied to a boundary value problem for a system of partial differential equations. For light rays, we are in the simpler situation just described: a light-like test particle in the field of a spherical body that undergoes a uniform translation in the frame $\mathrm{E}$, hence we need merely a summary of this scheme.

\subsubsection{Principle of the asymptotic PN approximation}\label{principlePNA}
Since Eq.~(\ref{field}) is a second-order hyperbolic equation, a gravitating system made of barotropic perfect fluids is defined \cite{A23} by the initial data for $f$ and $\partial_{T}f$ (for the scalar gravitational field), and by the initial data for the pressure $p$ and the absolute velocity $\mathbf{u} \equiv d\mathbf{x}/dT$ (for the matter fields). We want to describe gravitation in the {\it Newtonian limit} where: {\bf i}) the space metric $\Mat{g}$ (\ref{spacemetric}) becomes close to the Euclidean metric $\Mat{g}^0$ and the local time (\ref{localtime}) becomes close to the absolute time---both being true if $f \rightarrow  1$, and {\bf ii}) the matter fields become close to corresponding Newtonian fields. To do that, we use an exact similarity transformation that exists in Newtonian gravity \cite{FutaSchutz, A23}: If $p^{(1)}$, $\mathbf{u}^{(1)}$, $\rho^{(1)}=F^{(1)}(p^{(1)})$ (density), and $U_{N} ^{(1)}$ (Newtonian potential) obey the Euler-Newton equations, then, for any
$\lambda>0$, the fields 
\begin{equation}\label{similarity1}
p^{(\lambda)}(\mathbf{x},T)=\lambda^2p^{(1)}(\mathbf{x},\sqrt{\lambda}\ T), \quad
\rho^{(\lambda)}(\mathbf{x},T)=\lambda\rho^{(1)}(\mathbf{x},\sqrt{\lambda}\ T),
\end{equation}
\begin{equation}\label{similarity2}
U_{N}^{(\lambda)}(\mathbf{x},T)=\lambda U_{N}^{(1)}(\mathbf{x},\sqrt{\lambda}\ T), \quad
\mathbf{u}^{(\lambda)}(\mathbf{x},T)=\sqrt{\lambda}\ \mathbf{u}^{(1)}(\mathbf{x},\sqrt{\lambda}\ T),
\end {equation}
also obey these equations---provided the state equation for system $\mathrm{S}_\lambda$ is 
$F^{(\lambda)}(p^{(\lambda)}) = \lambda F^{(1)}(\lambda^{-2} p^{(\lambda)})$. This similarity
transformation defines the weak-field limit in Newton's theory itself: as $\lambda\rightarrow0$, the potential and the density in the bodies decrease like $\lambda$ (while the bodies keep the same size), the (orbital) velocities decrease like $\sqrt{\lambda}$, and accordingly the time scale increases like $1/\sqrt{\lambda}$. Therefore, to define the Newtonian limit in the scalar theory, we just apply the similarity transformation (\ref{similarity1})-(\ref{similarity2}) to the initial data for the gravitating system of interest. In doing so, $\rho$ becomes the proper rest-mass density and we substitute $V^{(\lambda)} = (c^2/2)(1-f^{(\lambda)})$ for the Newtonian potential $U_N^{(\lambda)}$. (The coefficient is needed to ensure that $V^{(\lambda)}\sim U_N^{(\lambda)}$ as $\lambda\rightarrow0$.) This defines a system ($\mathrm{S}_\lambda$) as the solution of this initial-value problem ($\mathrm{P}_\lambda$), and one expects that the solution fields admit expansions in powers of $\lambda$, whose dominant terms have the same orders in $\lambda$ as in Eqs.~(\ref{similarity1})-(\ref{similarity2}). Thus, we introduce a conceptual {\it family} of gravitating systems, with the real system of interest being assumed to correspond to a given, small value $\lambda_0$ of the parameter $\lambda $. It is then easy to check that, by adopting $[\mathrm{M}]_\lambda = \lambda[\mathrm{M}]$ and $[\mathrm{T}]_\lambda = [\mathrm{T}]/\sqrt{\lambda}$  as the new units for the system ($\mathrm{S}_\lambda$) (where $[\mathrm{M}]$ and $[\mathrm{T}]$ are the starting units of mass and time), all fields become order $\lambda^0$, and the small parameter $\lambda$ is proportional to $1/c^2$ (indeed $\lambda=(c_0/c)^2$, where $c_0$ is the velocity of light in the starting units). Hence, the derivation of the 1PN expansions and expanded equations is straightforward: {\it all fields} depend on the small parameter $1/c^2$, and one writes first-order expansions in this parameter. However, there is still a crucial point to be precised, that will also play a role in the discussion of gravitational radiation.

\subsubsection{The question of the time variable in the expansions} \label{timevariable}
The asymptotic expansions of the fields have first of all to be valid at fixed values of the time and space variables. Now, as we have just seen, the expansions are naturally formulated in units that depend on the small parameter $\lambda$. Hence, it is not equivalent to assume expansions with $x'^0 = T$ as the time variable, or with $x^0 = cT$, because the ratio $x'^0/x^0 = 1/c = \lambda^{1/2}/c_0$ depends on the small parameter. Note that, in the varying units, we have $T = \lambda^{1/2}T_0$ where $T_0$ is the "true" time ({\it i.e.} that measured in fixed units), hence $cT = c_0T_0$ is (proportional to) the true time. But the true orbital velocities in the system $\mathrm{S}_\lambda$ vary like $\lambda^{1/2}$, and the true orbital periods vary like $\lambda^{-1/2}$. Hence it is $T$, not $cT \propto T_0$, which remains nearly the same, as $\lambda$ is varied, for one orbital period of a given body in the {\it Newtonian limit}. Therefore, in this limit, thus for {\it post-Newtonian} expansions, one must take $x'^0 = T$ as the time variable, not $x^0 = cT$. The latter choice corresponds to {\it post-Minkowskian} (PM) expansions, with wave equations for the gravitational potentials, which are relevant in the study of gravitational radiation (see Section \ref{gravrad}). 

\subsubsection{Expansion of the metric and the field equation in the preferred frame}
The leading expansion is that of the scalar field $f = \beta^2$ (or rather, of the field $V= (c^2/2)(1-f)$, for this is really the field which is first-order expanded in $1/c^2$ below):
\begin{equation}\label{expans_f}
f = 1-2U/c^2-2A/c^4 + O(1/c^6)
\end{equation}
The space metric deduced from the Euclidean metric $\Mat{g}^0$ by Eq. (\ref{spacemetric}) is
\begin{equation}\label{expans_g}
g_{ij} = g^0_{ij}  + (2U/c^2)h^{(1)}_{ij} + O(1/c^4), \quad h^{(1)}_{ij} \equiv (U_{,i} U_{,j})/(g^{0kl}U_{,k}U_{,l})  
\end{equation}
(with $g^0_{ij} = g^{0ij}  = \delta_{ij}$  in Cartesian coordinates), and the space-time metric immediately follows by (\ref{spacetimemetric}):
\begin{equation}\label{gamma}
\gamma_{00} = f,\quad\gamma_{ij} = - g_{ij},\quad\gamma_{0i} = 0.  
\end{equation}
The mass-energy density $\sigma = T^{00}$ may be written in the form
\begin{equation}\label{expans_sigma}
\sigma = \sigma_0 + \sigma_1/c^2 + O(1/c^4),
\end{equation}
where $\sigma_0 = \rho_0$ is the conserved rest-mass density which is found at the first approximation. (Expanding, for a perfect fluid, the energy equation (\ref{energyconservation}), one finds that $\rho_0$ obeys the usual continuity equation; one also finds that mass is still conserved at the second (1PN) approximation.) The PN expansion of the field equation (\ref{field}) follows easily from Eqs. (\ref{expans_f}) and (\ref{expans_sigma}). Since the time variable is $x'^0 = T$ for PN expansions, the time-derivative term in (\ref{field}) occurs with a $c^{-2}$ factor as compared with the $\Delta f$ term, so that Poisson equations are obtained: 
\begin{equation}\label{expans_field_0}
\Delta U = -4\pi G \rho_0,				  
\end{equation}
\begin{equation}\label{expans_field_1}
\Delta A =  -4\pi G \sigma_1 + \partial^2 U/\partial T^2.	
\end{equation}

\subsection{Effects of a weak gravitational field on light rays} \label{lightrays}
\subsubsection{Post-Newtonian equations of motion for a photon in the preferred frame}
The application to the situation of a light-like test particle in the field of a spherical body that undergoes a uniform translation in the frame $\mathrm{E}$, with velocity $d\mathbf{x}/dT \equiv \mathbf{V}$, is not difficult \cite{A19}. For a photon, the velocity $u^i$ is $O(c)$. Using Eqs.~(\ref{expans_f}) and (\ref{expans_g}), one gets the 1PN expansion of the coordinate acceleration (\ref{coordinateacceleration}) for a photon. In Cartesian coordinates for metric $\Mat{g}^0$, this is
\begin{equation}\label{1PN_photon}
\frac{du^i}{dT}= U_{,i}- 2 \left(U_{,j}u^j\right)\frac{u^i}{c^2} - \left((Uh^{(1)}_{ij})_{,k}+(Uh^{(1)}_{ik})_{,j}-(Uh^{(1)}_{jk})_{,i}\right) \frac{u^j u^k}{c^2} +O\left(\frac{1}{c}\right).           
\end{equation}
The crucial point is that Eq.~(\ref{1PN_photon}), derived from Newton's second law, is nevertheless {\it undistinguishable from the PN expansion of the equation for null space-time geodesics}, as it is easily checked (\cite{A19}, after Eq.~(36)). Note that, in Eq.~(\ref{1PN_photon}), the Newtonian gravity acceleration $\mathbf{g}_0 \equiv \nabla_0 U$ (with components $U_{,i}$ in the Cartesian coordinates utilized) intervenes at the same order in $\lambda$ or $1/c^2$  as the other terms ({\it i.e.} the order zero, although it is really a second-approximation formula: the first approximation gives simply $du^i/dT = 0$). 
\subsubsection{Transition to the moving frame}
Let us denote by $\mathrm{E}_\mathbf{V}$ the frame that undergoes a pure translation, with the assumed uniform velocity $\mathbf{V}$ of the massive body, with respect to the ether frame $\mathrm{E}$. We may pass from $\mathrm{E}$ to $\mathrm{E}_\mathbf{V}$ by a Lorentz transformation of the flat space-time metric $\Mat{\gamma}^0$ (that one whose line element is $(ds^0)^2 = c^2 dT^2 - dx^i dx^i$ in Cartesian coordinates for the Euclidean space metric $\Mat{g}^0$). At the 1PN approximation, a photon follows a null geodesic of the physical space-time metric $\Mat{\gamma}$, and, by the Lorentz transformation, the components of this space-time tensor transform like a (twice covariant) tensor, of course. This gives an easy way to get the PN equations of motion for a photon in the frame $\mathrm{E}_\mathbf{V}$. Neglecting $O(1/c^4)$ terms, and except for an $O(1/c^3)$ term in the $\gamma'_{0i}$ components---such terms do not play any role in the 1PN expansion of the null geodesic equation \cite{Weinberg}---the components $\gamma'_{\mu\nu}$ after the Lorentz transform depend only on the ``Newtonian'' potential $U$, by just the same equations (\ref{expans_f})-(\ref{gamma}) as they do in the preferred frame. With the assumption of spherical symmetry, we have simply $U = GM_0/r$  with $M_0 \equiv \int \rho_0 dV$ : then, Eqs. (\ref{expans_f})-(\ref{gamma}) are just the 1PN expansion of Schwarzschild's exterior metric. Whereas this spherical potential is, in  general, not constant in the frame $\mathrm{E}$, it is indeed constant in the frame $\mathrm{E}_\mathbf{V}$ which moves with the spherical massive body ({\it e.g.} the Sun) that creates the relevant field, so that the geodesic equation is really that deduced from Schwarzschild's metric. We conclude that, to just the same level of approximation as in the 1PN approximation of GR, in particular neglecting the $\gamma_{0i}$ components of the metric, which are $O(1/c^3)$, the present theory predicts exactly the same gravitational effects on light rays as the standard effects obtained in GR with the Schwarzschild metric---and this is true accounting for the preferred-frame effects, for they do not appear at this approximation. The details can be found in Ref. \cite{A19}.

\section{Gravitational radiation (isolated system)} \label{gravrad}
In GR and in the other relativistic theories of gravitation, most investigations of gravitational radiation consider the so-called ``weak-field approximation'', which consists of a linearization of the field equations. In particular, d'Alembert equations are obtained for the gravitational potentials, which is the basic reason why gravitational radiation then occurs in a tractable form. Obviously, this linearization can be done also in the present theory with its nonlinear wave-type equation (\ref{field}). 

\subsection{Asymptotic post-Minkowskian approximation} \label{PMA}
With the goal to develop mathematically clear approximation schemes, it seems desirable to find a definite asymptotic framework which would indeed lead to a linearization of the field equations while keeping the propagating character. The PN scheme summarized in Subsection \ref{PNA} does not fit with this aim, since it leads to the Poisson-type equations (\ref{expans_field_0}) and (\ref{expans_field_1}). As it has been noted, this is due to the fact that, for consistency with the Newtonian limit, $x'^0 = T$ must be taken as the time variable, in the varying units $[\mathrm{M}]_\lambda = \lambda[\mathrm{M}]$ and $[\mathrm{T}]_\lambda = [\mathrm{T}]/\sqrt{\lambda}$, in which $\sqrt{\lambda}\propto c^{-1}$ and all fields are order zero. It appears from $\S$\ref{timevariable} that, in order to have $x^0 = cT$ as the time variable for different expansions, also based on this change of units, one must abandon the assumption that the velocity is low, and assume instead that it is order zero (in the starting units).

\subsubsection{Principle of the asymptotic PM approximation} \label{principlePMA}
Thus, let us define a family ($\mathrm{S}'_\lambda$) of perfect-fluid gravitating systems by the following family of initial conditions (in fixed units): at the initial time,
\begin{equation}\label{PM_IC1}
p^{(\lambda)}(\mathbf{x})=\lambda^2p^{(1)}(\mathbf{x}), \quad
\rho^{(\lambda)}(\mathbf{x})=\lambda\rho^{(1)}(\mathbf{x}),
\end{equation}
\begin{equation}\label{PM_IC2}
V^{(\lambda)}(\mathbf{x})=\lambda V^{(1)}(\mathbf{x}), \quad\partial_TV^{(\lambda)}(\mathbf{x})=\lambda \partial _TV^{(1)}(\mathbf{x}),
\end {equation}
\begin{equation}\label{PM_IC3}
\mathbf{u}^{(\lambda)}(\mathbf{x})= \mathbf{u}^{(1)}(\mathbf{x}).
\end {equation}
(Recall that $V\equiv (c^2/2)(1-f)$.) This describes a limit in which the gravitational field becomes weak (due to a correspondingly diminishing density), but nevertheless the velocity $u=\abs{\mathbf{u}}$ does not necessarily become negligible with respect to $c$---or, more exactly, $u/c$ has some finite order of magnitude, which does not depend on $\lambda$. We assume that, in a relevant time interval, the solution fields remain of the same order in $\lambda$ as in Eqs.~(\ref{PM_IC1})-(\ref{PM_IC3}). Then, in the units $[\mathrm{M}]_\lambda = \lambda[\mathrm{M}]$ and $[\mathrm{T}]_\lambda = [\mathrm{T}]/\sqrt{\lambda}$, we have again $\sqrt{\lambda}\propto c^{-1}$, and all fields: $p$, $\rho$, $V$, $\partial V/\partial x^0$, {\it but the velocity} $\mathbf{u}$, are order zero: as to $\mathbf{u}$, it is in fact order $c$.

\subsubsection{Zero-order PM approximation}\label{0PMA}
Since all fields but $\mathbf{u}$ are again order zero in the varying units, their (first-order) expansions have obviously the same form [Eqs.~(\ref{expans_f})-(\ref{expans_sigma})] as for the PN expansion. However, due to the fact that now $\mathbf{u}$ is $O(c)$, the zero-order term $\sigma_0$ in the expansion (\ref{expans_sigma}) of the active mass-energy density $\sigma \equiv T^{00}$ is not equal any more to the zero-order term $\rho_0$ of the density of rest-mass in the preferred frame---as this is the case \cite{A23} for the PN approximation. Instead, $\sigma_0$ now contains kinetic energy in relativistic form:
\begin{equation}\label{sigma0}
\sigma_0 = \rho_0\left(\gamma_v\right)_0 = \rho_0\left(1-\mathbf{u}_0^2\right)^{-1/2}\equiv \tau ,
\end {equation}
where $\mathbf{u}_0$ is the coefficient of $c$ in the PM expansion of $\mathbf{u}$: $\mathbf{u}= \mathbf{u}_0 c + O(c^{-1})$.\\

Because $x^0 = cT$ is now the time variable, the time-derivative term in Eq.~(\ref{field}) for the scalar field $f$ now does not contain the $c^{-2}$ factor, hence it is expanded to a ``flat space-time wave equation'' with d'Alembert operator:
\begin{equation}\label{expans_field_PM0}
\square U \equiv U_{,0,0}-\Delta U = 4\pi G \tau \qquad (x^0 = cT).				  
\end{equation}
The dynamical equations (\ref{flat_0})-(\ref{flat_i}), as applied to a perfect fluid, take the exact form \cite{A23}:
\begin{equation}\label{fluid_T}
\partial_T \left(\psi f\right) + \partial_j \left(\psi f u^j\right) = \left(\sigma /2 \right)\partial_T f + c^{-2}\partial_T p,
\end{equation}
\begin{equation}\label{fluid_i}
\partial_T \left(\psi u^i \right)	 + \partial_j \left(\psi u^i u^j \right) + \Gamma^i_{jk}\psi u^j u^k + \mathsf{t}^i_k \psi u^k = \psi f \mathsf{g}^i - g^{ij}p_{,j},  
\end{equation}
where
\begin{equation}\label{psi_and_t}
\psi \equiv \sigma + \frac{p}{c^2 f}, \qquad \mathsf{t}^i_k \equiv \frac{1}{2}g^{ij}\partial_T g_{jk}.
\end{equation}
When the PN approximation is used (time variable $T$, all fields of order zero), the zero-order expansion of these equations gives \cite{A23} the continuity equation and the classical fluid-dynamics equation in the presence of the gravity acceleration $U_{,i}$. (This equation can be put in the form of Euler's equation using the continuity equation.) In contrast, with the PM approximation (time variable $x^0=cT$, and $\mathbf{u}$ of order $c$), we get a ``continuity equation'' which involves the density $\tau $ (\ref{sigma0}) instead of the 0-order rest-mass density:
\begin{equation}\label{fluid_PM0_0}
\partial_0 \tau  + \partial _j (\tau u_0^j) = 0 \qquad (x^0=cT),
\end{equation}
plus an equation meaning that the principal part of the acceleration is {\it zero}:
\begin{equation}\label{fluid_PM0_i}
    \partial_0 (\tau u_0^i) + \partial _j (\tau u_0^i u_0^j) = 0.
\end{equation}
It should not surprise us that, when we do not impose the Newtonian limit, we do not get Newtonian equations indeed! The ``asymptotic'' scheme of PM approximation that we have introduced in $\S$\ref{principlePMA} (seemingly for the first time) is more general than the asymptotic PN scheme recalled in $\S$\ref{principlePNA}, in the sense that there is just more freedom left to the magnitude of the velocity. Thus, what happens here is that, due to the fact that $\mathbf{u}$ is order $c$, the acceleration $U_{,i}$ plays a role only at the following order (1PM approximation)---but the latter seems to be rather complicated. However, the zero-order PM approximation is enough to assess the principal part of the radiative energy loss.

\subsection{0PM approximation of the radiative energy loss}
Considering an isolated self-gravitating system, assumed to be suitable for the PM scheme above (thus the gravitational field is weak, but the orbital velocities are not {\it necessarily} small), we wish to investigate the asymptotic behaviour, as $R_0 \rightarrow \infty$, of the energy variation (\ref{energyvariation}) in the ball $\Omega:\abs{\mathbf{X}}\leq R_0$. More exactly, in the framework of the PM approximation, we investigate the limit, as $R_0 \rightarrow \infty$, of the {\it principal part} of $dE/dT$ as $\lambda \rightarrow 0$. Thus, we evaluate the flux $\Phi$ (\ref{energyvariation}) at the zero-order PM (0PM) approximation. We have then from (\ref{g_grav}) and (\ref{expans_f}):
\begin{equation}\label{PM0_flux}
\Phi = \Phi_0\left(1+O(c^{-2})\right), \quad \Phi_0\equiv \frac{-1}{4\pi G} \int_{\partial \Omega} \partial_T U\,U_{,k}n^k\,dS. 
\end{equation}
\subsubsection{Expansion of the retarded potential}
The relevant solution of Eq.~(\ref{expans_field_PM0}) is the retarded potential:
\begin{equation}\label{U_ret}
U(\mathbf{X},x^0)= \int \frac{G\tau (\mathbf{x},x^0-\abs{\mathbf{X-x}})}{\abs{\mathbf{X-x}}}\,dV(\mathbf{x}). 
\end{equation}
(The integral extends over the whole ``retarded system'', which is spatially bounded as is the system itself. Moreover, the point $\mathbf{X}$ shall be at a large distance from the retarded system.) The expansion of $U$ as $R_0 \equiv \abs{\mathbf{X}}\rightarrow \infty$ is got by a standard procedure (see {\it e.g.} Stephani \cite{Stephani}, pp. 125-126), which we try to make mathematically explicit---though without giving a convergence proof. One writes first the Taylor series expansion at $x^0-R_0$ of the retarded density: 
\begin{equation}\label{Taylor_tau_ret}
\tau(\mathbf{x},x^0-R) =  \tau(\mathbf{x},x^0- R_0) +  \delta R \, \partial_0\tau(\mathbf{x},x^0-R_0)+\frac{{\delta R}^2}{2!}\partial^2_0\tau(\mathbf{x},x^0- R_0)+ ...,
\end{equation}
where
\begin{equation}
R \equiv \abs{\mathbf{X-x}}, \quad R_0 \equiv \abs{\mathbf{X}},  \quad \delta R \equiv R_0-R.
\end{equation}
Inserting into (\ref{Taylor_tau_ret}) the first-order Taylor expansion of $R$ at $\mathbf{X}$:
\begin{equation}
\delta R = \frac{X^ix^i}{R_0} + O\left(\frac{1}{R_0}\right),
\end{equation}
one gets:
\begin{eqnarray}\label{expans_tau_ret}
\tau(\mathbf{x},x^0-R) & =  &\tau(\mathbf{x},x^0- R_0) +  \partial_0\tau(\mathbf{x},x^0-R_0)\left(\frac{X^ix^i}{R_0}\right) \nonumber \\
& + & \frac{\partial^2_0\tau(\mathbf{x},x^0- R_0)}{2!}\left(\frac{X^iX^jx^ix^j}{R_0^2}\right)+... + O\left(\frac{1}{R_0}\right),
\end{eqnarray}
where the points of suspension indicate omitted terms of order $R_0^0$ coming from the remainder in the infinite series (\ref{Taylor_tau_ret}). Using then the zero-order expansion of $1/\abs{\mathbf{X-x}}$ at $\mathbf{X}$:
\begin{equation}\label{expans_1/R}
\frac{1}{\abs{\mathbf{X-x}}}=\frac{1}{R_0} + O\left(\frac{1}{R_0^2}\right),
\end{equation}
together with (\ref{expans_tau_ret}), and integrating, gives the expansion of $U$ up to $1/R_0$:
\begin{equation}\label{expans_U_ret}
\frac{U(\mathbf{X},T)}{G}=\frac{M}{R_0} + \frac{X^i\dot{d}_i}{cR_0^2} + \frac{X^iX^j}{2c^2R_0^3}\ddot{J}_{ij} +...+ O\left(\frac{1}{R_0^2}\right),
\end{equation}
where overdot means partial derivative with respect to $T$, the points of suspension indicate omitted terms of order $1/R_0$ coming from time derivatives of order $\geq 3$ in (\ref{Taylor_tau_ret}), and with
\begin{eqnarray}\label{def_M_d_J}
M\equiv \int \tau dV, & \quad d_i(\mathbf{X},T) \equiv \int x^i\tau(\mathbf{x},cT-\abs{\mathbf{X}}) dV(\mathbf{x}),\\ & J_{ij}(\mathbf{X},T) \equiv \int x^ix^j\tau(\mathbf{x},cT-\abs{\mathbf{X}}) dV(\mathbf{x}). 
\end{eqnarray}
\subsubsection{Limiting value of the flux}
When taking the partial space-derivatives, we get from the definition (\ref{def_M_d_J}):
\begin{equation}\label{drond_di_drond_Xk}
\frac{\partial d_i}{\partial X^k}=\frac{-X^k}{cR_0}\dot{d}_i, 
\end{equation}
and the like for $J_{ij}$. We obtain therefore from (\ref{expans_U_ret}):
\begin{equation}\label{flux_integrand}
\frac{\partial_TU\partial_kU}{G^2} = \left[\frac{X^i\ddot{d}_i}{cR_0^2} + \frac{X^iX^j}{2c^2R_0^3}\dot{\ddot{J}}_{ij} \right]\left[\frac{-X^lX^k\ddot{d}_l}{c^2R_0^3} + \frac{-X^lX^mX^k}{2c^3R_0^4}\dot{\ddot{J}}_{lm} \right]+ O\left(\frac{1}{R_0^3}\right).
\end{equation}
(We have $dM/dT=0$ by Eq.~(\ref{fluid_PM0_0}).) But it is also true that
\begin{equation}\label{canceldipole}
\ddot{d}_i=0. 
\end{equation}
Indeed, the retardation being uniform in the definition (\ref{def_M_d_J}), this is a standard exact consequence of Eqs.~(\ref{fluid_PM0_0}) and (\ref{fluid_PM0_i}), for the continuum with density $\tau \equiv \sigma_0$ and that moves with velocity $c\mathbf{u}_0$---the acceleration field of this continuum being zero in view of Eq.~(\ref{fluid_PM0_i}). On the other hand, in the next subsection we shall consider the same physical system at the {\it Newtonian} approximation, thus equating $\tau$ in the definition (\ref{def_M_d_J}) with the zero-order PN rest-mass density (which is consistent with (\ref{sigma0}) insofar as the actual velocity $\mathbf{u}$ is in fact negligible with respect to $c$). But at this approximation also it is true that $\ddot{d}_i=0$, because the total momentum of an isolated Newtonian system is conserved.\\

Thus, there is no dipole term in fact. From (\ref{PM0_flux}), (\ref{flux_integrand})
and (\ref{canceldipole}), we get:
\begin{equation}\label{limit_flux1}
-4\pi G \lim_{R_0\rightarrow \infty}\Phi_0 = \frac{-G^2}{4c^5}\int \left(\dot{\ddot{J}}_{ij}n^i n^j\right)^2 d\omega, 
\end{equation}
where $d\omega$ is the solid angle element and $n^i\equiv X^i/R_0$. (Note that tensor 
$\dot{\ddot{\mathbf{J}}}$ is constant on the sphere $\abs{\mathbf{X}}=R_0$.) Using the integration formula for $n^i n^j n^k n^l$ \cite{Fock59} (Eq.~(90.20)), it follows that
\begin{equation}\label{limit_flux2}
(\Phi_0)_{\mathrm{lim}} = \frac{G}{60c^5} \left(2\,\mathrm{tr}\,\dot{\ddot{\mathbf{J}}}^2 + \mathrm{tr}^2 \dot{\ddot{\mathbf{J}}}\right).
\end{equation}

\subsection{Peters-Mathews coefficients for a binary system}\label{Peters-Mathews}
The energy loss $\dot{E}=-(\Phi_0)_{\mathrm{lim}}$ is the principal part of the variation of the total energy of our isolated weakly-gravitating system, due to its emission of gravitational waves. As it was already mentioned at several places in this Section, this unique system may consistently be treated in both the PM approximation and the PN one. The (more general) PM approximation has allowed us to calculate an approximate value of the energy loss, and now we are going to use the 0PN (Newtonian) approximation to see how the energy loss changes the relative motion in the first approximation. To do this means assuming: i) that the 0PN approximation of the energy of the system (the Newtonian energy, which is an exact constant in the framework of the 0PN approximation), actually varies as the exact energy of the system---and ii) that the latter variation is well approximated by the 0PM calculation. \footnote{\,We note that the same two assumptions have to be done when one applies the quadrupole formula of GR, and that assumption i) may seem strong. {\it A priori}, one might expect that the energy loss found at the 0PM approximation would affect also the higher-order terms (possibly even {\it only} the higher-order terms!) in the PN energy. Of course, the Newtonian energy dominates over the PN corrections to it. Assumption i) is valid if this remains true for the corresponding {\it rates}, which should be true if the PN expansions are uniform with respect to time.} 
We consider more particularly a binary system, made of two massive bodies that are far enough from each other, so that tidal interactions are negligible and each body has a rigid motion. We assume, moreover, that the self-rotation of each body is not modified by the energy loss, which thus affects merely the relative motion of the two bodies, whose motion may then be analysed as a Newtonian 2-body problem for two point masses $m_1$ and $m_2$, located at $\mathbf{x}_1(T)$ and $\mathbf{x}_2(T)$. However, since $\dot{E}$ has been evaluated in the preferred frame $\mathrm{E}$, this Newtonian problem is formulated in that frame (which is indeed an inertial frame for the 0PN approximation of the theory). Setting
\begin{equation}\label{def_x_u}
\mathbf{x}\equiv \mathbf{x}_1-\mathbf{x}_2, \qquad \mathbf{u}\equiv \dot{\mathbf{x}},
\end{equation}
we have thus
\begin{equation}\label{u_dot}
\dot{\mathbf{u}} = -k\mathbf{x}/r^3, \quad k\equiv G(m_1+m_2), \quad r\equiv \abs{\mathbf{x}}.
\end{equation}
The quadrupole tensor is then:
\begin{equation}\label{binaryJ}
\mathbf{J}= m_1 \mathbf{x}_1\otimes \mathbf{x}_1 + m_2 \mathbf{x}_2 \otimes \mathbf{x}_2. 
\end{equation}
Introducing the mass center $\mathbf{a}(T)$, such that $m_1\mathbf{x}_1 +m_2\mathbf{x}_2=M\mathbf{a}$, where $M\equiv m_1+m_2$, we obtain
\begin{equation}\label{J-Jtilda}
\mathbf{J}= \tilde{\mathbf{J}} + M\mathbf{a}\otimes \mathbf{a}, \quad \tilde{\mathbf{J}} \equiv m_1 (\mathbf{x}_1-\mathbf{a})\otimes (\mathbf{x}_1-\mathbf{a})+ m_2 (\mathbf{x}_2-\mathbf{a})\otimes (\mathbf{x}_2-\mathbf{a}). 
\end{equation}
Since the system is isolated, the velocity $\mathbf{V}\equiv \dot{\mathbf{a}}$ is a constant. It follows that 
\begin{equation}\label{Jtilda3dot}
\dot{\ddot{\mathbf{J}}} = \dot{\ddot{\tilde{\mathbf{J}}}}. 
\end{equation}
We get from (\ref{def_x_u}):
\begin{equation}\label{Jtilda}
\tilde{\mathbf{J}} = \mu \mathbf{x}\otimes \mathbf{x}, \quad \mu \equiv m_1m_2/M.
\end{equation}
Three time differentiations of (\ref{Jtilda}) give us with (\ref{u_dot}) and (\ref{Jtilda3dot}):
\begin{equation}\label{binaryJ3dot}
\dot{\ddot{\mathbf{J}}} = \frac{2k\mu}{r^2}\left[3(\mathbf{u.n})\,\mathbf{n}\otimes \mathbf{n}-2(\mathbf{n}\otimes \mathbf{u}+\mathbf{u}\otimes \mathbf{n})\right], \quad \mathbf{n}\equiv \frac{\mathbf{x}}{r},
\end{equation}
whence follows:
\begin{equation}\label{trJ3dot}
\mathrm{tr}\, \dot{\ddot{\mathbf{J}}} = -\frac{2k\mu}{r^2}\mathbf{u.n}, \quad \mathrm{tr}\, \dot{\ddot{\mathbf{J}}}^2 = 
\frac{4k^2\mu^2}{r^4}\left[8\,\mathbf{u}^2-7\,\left(\mathbf{u.n}\right)^2\right]. 
\end{equation}
Using Eq.~(\ref{limit_flux2}), this leads finally to
\begin{equation}\label{binaryEdot}
\dot{E}=-\left(\Phi_0\right)_\mathrm{lim} = -\frac{8}{15}\,\frac{G}{c^5} \,\frac{k^2\mu^2}{r^4}\left[2\,\mathbf{u}^2-\frac{3}{4}\left(\mathbf{u.n}\right)^2\right], 
\end{equation}
{\it i.e.} the so-called Peters-Mathews coefficients (see {\it e.g.} Will \cite{Will}, Eq.~(10.84)):
\begin{equation}\label{PetersMathews}
k_1 = 2,\quad k_2=\frac{3}{4}, \quad k_\mathrm{dipole}=0. 
\end{equation}
There is no average to be performed over one binary's orbit in Eq.~(\ref{limit_flux2}) nor in Eq.~(\ref{binaryEdot})---in contrast with what is often found in the literature. If we start from the ``quadrupole formula of GR'' (the equivalent, for GR in harmonic coordinates and for the RTG, of Eq.~(\ref{limit_flux2}) here: {\it cf.} Ref. \cite{L&L}, Eq.~(110,16), and Ref. \cite{Logunov89}, Eq.~(15.71), where there is indeed no time average):
\begin{equation}\label{4poleGR}
 \dot{E} = -\frac{G}{45c^5} \mathrm{tr}\left(\dot{\ddot{\mathbf{D}}}^2\right) =-\frac{G}{5c^5} \left(\mathrm{tr}\,\dot{\ddot{\mathbf{J}}}^2 -\frac{1}{3}\,\mathrm{tr}^2 \dot{\ddot{\mathbf{J}}} \right), \quad \mathbf{D}\equiv 3\left[\mathbf{J}-\frac{1}{3}\,\left(\mathrm{tr}\,\mathbf{J}\right)\mathbf{I}\right],
\end{equation}
and apply Eq.~(\ref{trJ3dot}), we find
\begin{equation}\label{binaryEdotGR}
\dot{E}= -\frac{8}{15}\,\frac{G}{c^5} \,\frac{k^2\mu^2}{r^4}\left[12\,\mathbf{u}^2-11\left(\mathbf{u.n}\right)^2\right], 
\end{equation}
which is indeed the Peters-Mathews formula of GR in harmonic coordinates \cite{Will} and of the RTG \cite{Logunov89}, without any time average (in contrast with Ref.~\cite{Will}, Eq.~(10.80)). The time-average may appear physically justified, for one expects that the time-period of the gravitational waves should be of the order of the orbital period.  But, as we have outlined at the beginning of this subsection, it seems in fact reasonable to admit that the rate of the radiative energy loss equals the decrease-rate of the Newtonian energy. Once this equality is admitted, it implies a ``Peters-Mathews formula'' without any time-average.

\section{Conclusion}
We have given a nearly self-consistent summary of the ``scalar ether-theory'': Eqs.~(\ref{g_grav}) for the gravity acceleration, (\ref{spacemetric}) and (\ref{spacetimemetric}) for the metric, and (\ref{field}) for the scalar field, can be taken axiomatically. We have more detailed its dynamics (Eq.~(\ref{Newtonlaw}) for a test particle, Eq.~(\ref{continuum}) for a continuum). It should be noted that this dynamics is independent of the assumptions made as to the form of the metric. A detailed motivation for the theory, together with a summary of its construction, can be found in Refs.~\cite{A28} or \cite{B13}.\\

This scalar theory gives the same predictions for light rays as the standard predictions deduced from Schwarzschild's metric in GR. This is because the theory predicts Schwarzschild's metric (Eq.~(\ref{SSSmetric})), and because it turns out that the preferred-frame effects do not play any role at the first post-Newtonian (1PN) approximation {\it for photons}. In turn, the fact that this central metric of GR also does occur in this scalar theory, is related to its preferred-frame character, which allows to postulate gravitational time-dilation and space-contraction in a way that is not space-time covariant.\\

To investigate gravitational radiation, we have introduced an ``asymptotic'' form of the post-Minkowskian (PM) approximation, based on an explicit family of initial-value problems for a perfect-fluid system (Eqs.~(\ref{PM_IC1})-(\ref{PM_IC3})). This seems to be new. We did not prove the existence of the solutions, nor did we prove the truly asymptotic character of the expansions, but we consider it likely that both could be proved (in suitable function spaces). The important mathematical work \cite{DamourSchmidt}, that is devoted to GR (whose gravitational field equations are much more complex), does give such justifications, but it is restricted to the {\it vacuum} Einstein equations. As it was the case for our ``asymptotic'' PN scheme \cite{A23}, the present PM scheme justifies, by invoking a change of units, to take $1/c^2$ as the small parameter, while considering that (nearly) all fields are order zero---but {\it all} fields depend on the small parameter and hence must be expanded. The difference between these PN and PM schemes is that, in the latter, the time variable is $x^0=cT$, not $T$, and the velocity is $O(c)$ only. But the latter is true in the changed units and in an asymptotic sense ({\it i.e.} for the conceptual one-parameter family of gravitational systems), physically it means that the velocity field {\it can} be non-negligible with respect to $c$, but it does not {\it have to be} so. Thus this PM scheme is just {\it more general} than the corresponding PN scheme. This is the reason why the energy rate calculated in the 0PM approximation can be used to investigate the time evolution of the system, when the latter is considered at the PN approximation. Thus, within this framework, it seems that one can reasonably assume that it is not undispensable, in a first approximation of the ``radiation reaction effects'', to make a complete analysis of the relativistic 2-body problem up to the order $1/c^5$ or higher. Such a complete analysis has been investigated for GR ({\it cf.} Damour \cite{Damour87} and references therein), but it is very complex. \\

Using the PM approximation, we have shown that, in this scalar theory with a preferred reference frame, there is a flux of gravitational energy which emanates from an isolated weakly-gravitating system, and which goes to infinity; that this limit flux is positive, Eq.~(\ref{limit_flux2}) (hence the system {\it loses} energy); and that it starts only from quadrupole terms, the monopole and dipole terms being zero. When equated to the energy rate of a binary system in the Newtonian approximation, this energy loss gives a Peters-Mathews formula with well-ordered positive coefficients, again without any dipole (Eq.~(\ref{binaryEdot})). Since these coefficients differ from those in GR or in the RTG, a different value of the energy loss would be obtained (albeit of the same order of magnitude), if one would apply Eq.~(\ref{binaryEdot}) while otherwise keeping the parameters of the 2-body system at the values found \cite{TaylorWeisberg} from fitting the pulse data with a ``timing model'' based on GR. However, there is essentially only one input data for the timing model, namely the list of the time intervals between the successive pulses emitted by the pulsar (in short ``the pulse data"), and if the theory is changed, then, after fitting the pulse data by a timing model based on the new theory, different values will be found also for the parameters of the 2-body system, {\it e.g.} for the eccentricity and for the masses of the two bodies. Therefore, the author feels it plausible that the pulse data of binary pulsars might be nicely fitted by building a timing model entirely based on this theory, in the same way as they have been fitted in GR \cite{TaylorWeisberg}. What can only be stated for now, is that the theory predicts an energy loss with the same order of magnitude to that which is found in GR and which fits with the observations. As emphasized by Will~\cite{Will}, most alternative theories do not predict the same order of magnitude as in GR for the energy rate, and not rarely they even predict the wrong sign---{\it e.g.} this is the case for Rosen's bimetric theory~\cite{Rosen}, that looks much closer to GR than looks the present scalar theory.\\

{\large{\bf Acknowledgment}}

I am grateful to the referee for his suggestion to calculate the Peters-Mathews coefficients and for his remarks on the result, in particular for his calculation showing that, for the binary pulsar PSR1913+16, the energy loss predicted by the scalar theory would be some six times less than in GR, if one would take the other parameters of the system as obtained from fitting the pulse data by the timing model based on GR.


\bibliographystyle{amsplain}

\end{document}